\documentclass[12pt]{article}
\usepackage{graphicx}
\evensidemargin  \oddsidemargin
\usepackage{amsmath}
\usepackage{amssymb}
\begin{document}

\noindent 
{\bf A Numerical Study of a Simple Stochastic/Deterministic Model 
of Cycle-to-Cycle Combustion Fluctuations in Spark Ignition 
Engines}

\vspace{1cm}

\noindent 
G. LITAK \\
{\it Department of Applied Mechanics, 
Technical University of Lublin, Nabystrzycka 36, \\
PL-20-618 Lublin, Poland} \\ \\

\noindent
M. WENDEKER \\
M. KRUPA \\
{\it Department of Internal Combustion Engines, 
Technical University of Lublin, \\  Nabystrzycka 36,
PL-20-618 Lublin, Poland} \\ \\

\noindent
J. CZARNIGOWSKI \\
{\it Department of Machine Construction, 
Technical University of Lublin, Nabystrzycka 36,
PL-20-618 Lublin, Poland}

\vspace{1cm}

\noindent
We examine a simple, fuel-air, model of 
combustion
in spark ignition (si) engine with indirect injection.
In our two fluid model, variations of fuel mass burned in cycle 
sequences appear
due to stochastic 
fluctuations of a fuel feed amount. We have shown 
that a small amplitude of these fluctuations affects  
considerably the stability of a 
combustion process strongly depending on the quality of air-fuel mixture. 
The largest influence was found in the limit of a lean combustion.   
The possible effect of nonlinearities in the 
combustion process has been also 
discussed.

\vspace{0.5cm}



\section{Introduction}

\noindent Cyclic combustion variability,  found in 19th century 
by Clerk (1886) in
all spark
ignition (si) engines, has attracted a great interest of researchers during 
last 
years (Heywood 1988, Daily 1988, Foakes {\em et al.} 1993,  Chew {\em et al.} 1994, Hu 1996, Daw {\em et al.} 
1996, 1998, 2000, Letellier {\em et al.} 1997,
Green Jr. {\em et al.} 1999, Rocha-Martinez {\em et al.} 2002, Wendeker
{\em et al.} 2003, 2004, Kami\'nski {\em et al.} 2004). 
Its elimination 
 would give 10\%
increase
in the power output  of the engine.  
The main sources of cyclic variability were
classified by Heywood
(1988) as
the aerodynamics in the cylinder during
combustion, the amount of fuel, air and recycled exhaust
gases supplied to the cylinder and a
mixture composition near the spark plug.

The key source of their existence may be associated with either 
stochastic disturbances 
(Roberts {\em et al.} 1997,
Wendeker {\em et al.} 2000) 
or 
nonlinear dynamics   
(Daw {\em et al} 1996, 1998)
 of 
the combustion process. 
Daw {\it et al.} 
(1996, 1998)
  and more recently
Wendeker {\it et al.} 
(2003, 2004)
have done  the nonlinear analysis of the experimental 
data of such a process.

Various attempts have been done to explain this phenomenon 
Shen {\em et al.} (1996) modelled 
a kernel and a front of flame in the cylinder. Their motions and interactions with cylinder chamber walls 
influenced the region of combustion leading to  turbulent behaviour.
Hu (1996) developed a phenomenological combustion model and examined the system answer 
on small variations of different combustion parameters as changes in fuel-air mixture. Stochastic models 
of internal combustion  basing on residual gases where considered by Daw {\em et al.} (1996, 1998, 2000),
Roberts {\em et al.} (1997),
Green Jr. {\em et al.} (1999), and recently  by Rocha-Martinez {\em et al.}  (2002), who examined multi-input
combustion models.  
Daw and collaborators described the combustion process using 
a recurrence model with nonlinear combustion efficiency
and invented   
symbolic analysis for combustion stability. 

In this paper follow these investigations with  a new  model 
assuming that 
variations of a fresh fuel  amount is the 
most 
common source of instability in indirect injection.

The present paper is organised as follows. After the introduction in the 
present 
section we define the model by a set of difference equations in the next 
section (Sec. 2). 
 This model, in deterministic and stochastic forms, will be applied in 
Sec. 
3, where we analyse the oscillations of burned mass. Finally we 
derive conclusions and remarks in Sec. 4.  

\section{Two fluid model of fuel-air mixture combustion}

\noindent Starting from fuel-air mixture
we define the time evolution of the corresponding amounts.
Namely, we will follow
the time histories of the masses of fuel $m_f$, and air $m_a$.

\begin{table}
\caption{\label{tableone} Constants and variables of the model.
}
\vspace{1cm}
\hspace{2cm}
\begin{tabular}{c|c}
 \hline
stoichiometric coefficient  & $1/s=14.63$
 \\
residual gas fraction    & $\alpha=0.08$, 0.16  \\
air mass in a cylinder  & $m_a$    \\
fuel mass in a cylinder  & $m_f$  \\
fresh air amount & $\delta m_a$   \\
fresh fuel amount & $\delta m_f$  \\
air/fuel ratio & $r=m_a/m_f$  \\
burned fuel mass & $\Delta m_f$  \\
combusted air mass & $\Delta m_a$  \\
air/fuel equivalence ratio & $\lambda$  \\ 
random number generator & $ N(0,1,i)$  \\
mean value \\ of fresh fuel amount & $\delta m_{fo}$  \\
standard deviation \\of fresh fuel amount & $\gamma=\sigma_{mf}$  \\
standard deviation \\ of  the equivalence ratio & $\sigma_{\lambda}$  \\
\hline
\end{tabular}
\vspace{1cm}
\end{table}

\begin{figure}
\begin{center}
\includegraphics[scale=0.4,angle=-90]{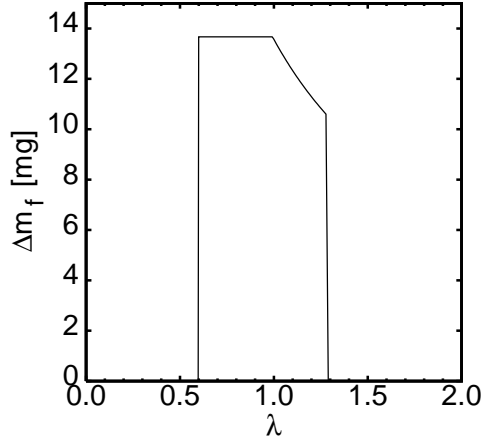}
\caption{The  combustion curve $\Delta m (\lambda)$ for the constant  
fresh air feed
$\delta 
m_a=200$ mg.}
\end{center}
\end{figure}

Firstly, we assume the initial value of  $m_a(i)$, $m_f(i)$ 
and automatically their ratio
$r(i)$:
\begin{equation}
r(i) =\frac{m_a(i)}{m_f(i)}
\end{equation}
for  $i=0$.

Secondly, depending on parameter $r$ with reference to a stoichiometric 
constant $s$ we have two possible cases:
 fuel and air deficit, respectively.
For a deterministic model, the first case lead to  
\begin{equation}
r(i) > 1/s
\end{equation}
we calculate next masses using following difference equations:
\begin{eqnarray}
m_f (i+1) &=& \alpha \left(m_f(i) -\frac{1}{s}
m_a (i) \right) + \delta m_f
\nonumber \\
m_a (i+1) &=&  \delta m_f,
\end{eqnarray}
where $\alpha$ is the residual gas fraction of the engine,
 $\delta m_f$ and  $\delta m_a$ denotes fresh fuel and air amounts added 
in each
combustion cycle $i$. 
In the opposite (to Eq. 2)  case
\begin{equation}
r(i) < 1/s
\end{equation}
we use the different formula
\begin{eqnarray}
m_f (i+1) &=&  \delta m_f
\nonumber \\
m_a (i+1) &=& \alpha \left(m_a(i)- s m_f(i) \right) + \delta m_a
\end{eqnarray}

Note that variables $m_a$ and $m_f$ are the minimal set of our interest.
From the above equations  one can easily calculate other interesting 
quantities as the combusted masses of 
fuel $\Delta m_f(i)$ and air $\Delta m_a(i)$ and air-fuel equivalence 
ratio before each combustion event $i$:
\begin{equation}
\lambda \approx s\frac{m_a(i)+\alpha\Delta 
m_a(i-1)}{m_f(i)+\alpha\Delta 
m_f(i-1)}  
\end{equation}
Basing on experimental results we use the additional necessary condition 
(Kowalewicz 1984)
of
combustion process
\begin{equation}
0.6 < \lambda < 1.3 . 
\end{equation}
For better clarity our notations of system parameters: constants and 
variables are 
summarised in Tab. 1.

Basing on the relations (Eqs. 1-7) we plotted the combustion 
curve for the assumed constant fresh air feed $\delta m_a=200$ mg.
This value will be used for all simulations throughout this paper.

\begin{figure}
 \begin{center}
\includegraphics[scale=0.3,angle=-90]{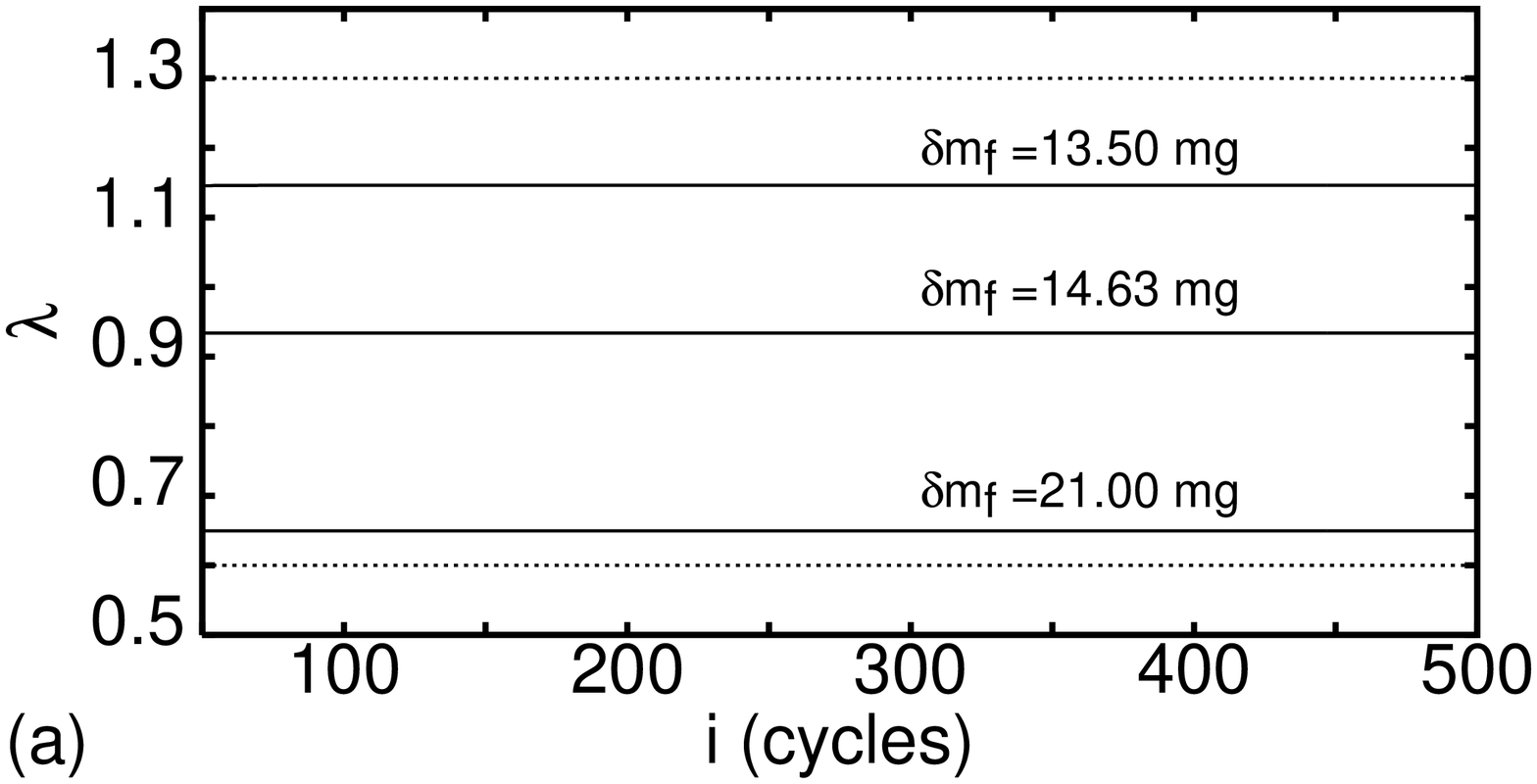}
\includegraphics[scale=0.3,angle=-90]{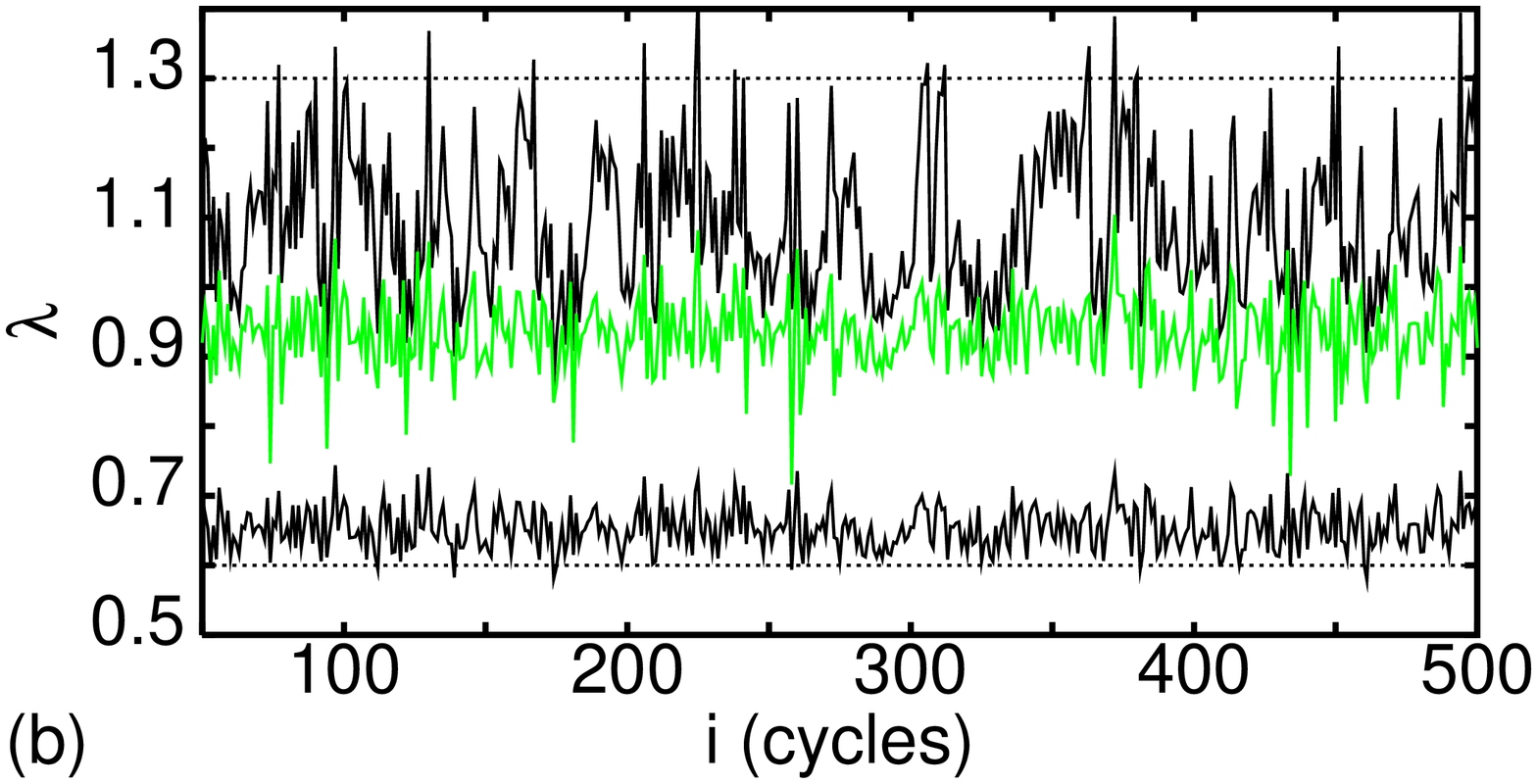}
\caption{The dependence of $\lambda$ on sequential cycles $i$ 
for deterministic (a) and 
stochastic (b) processes. The fresh 
air amount was assumed to be $\delta m_a=200$ mg while the fresh
fuel amount varies: $\delta m_f=13.50$ mg , 14.63 mg, 21.00 
mg starting from the top
curve,
respectively.}
 \end{center}
\end{figure}

Finally, in the case of stochastic injection,  instead of 
constant $\delta m_f(i)$ (Eqs. 3 and 5) (for each cycle $i$) we 
introduce
its mean value  $\delta m_{fo}={\rm const.}$, while $\delta m_f$ in the 
following way:
\begin{equation}
\delta m_f(i)=\delta m_{fo}+ \gamma N(0,1,i),
\end{equation}
where $N(0,1,i)$ represents random number generator giving a sequence $i$ 
of numbers with  a 
unit-standard deviation of normal 
(Gaussian) 
distribution and the nodal mean. The scaling factor $\gamma=\sigma_{mf}$ 
corresponds to the 
mean standard deviation of the fuel injection amount.
The cyclic variation of $\delta m_f(i)$ can be associated with such 
phenomena as  fuel vaporisation and  fuel-injector variations.

\section{Oscillations of burned fuel mass}

\noindent Here we describe the results of simulations. Using Eqs. 1-8
we have performed recursive calculations for deterministic and 
stochastic conditions and obtained time histories of 
various system parameters:
$m_f$, $m_a$, $\Delta m_f$, $\Delta m_a$ and $\lambda$.
The results for $\lambda$ are shown in Fig. 2. The upper panel (Fig. 2a), 
corresponding to deterministic combustion for three different values of  
fuel 
injection parameter $\delta m_{f}$, 
shows $\lambda$ as straight lines versus cycle $i$, while the lower 
(Fig. 2b) one
reflects the variations of $\lambda$ in stochastic conditions. 
The order of curves appearing in the Fig. 2b  is the same as in Fig. 2a
stating from the smallest value of considered fuel injection amounts from 
the top.  
To get a more clear insight of random
fuel injection on the engine dynamics
, in our stochastic simulations, we assumed  
standard deviation of its mean value
$\gamma=\sigma_{mf}= 0.1~ \delta m_{fo}$
to be high enough.
In following  calculations it was equal to 10\%. 
The obtained results clearly indicate that the  fluctuations of $\lambda$
are growing with larger $\lambda$. This can be also found by analytical 
evaluation of Eq. 6. It is not difficult to check that
\begin{equation}
\sigma_{\lambda} \sim \lambda^2 \sigma_{mf}.
\end{equation}

\begin{figure}
 \begin{center}
\includegraphics[scale=0.3,angle=-90]{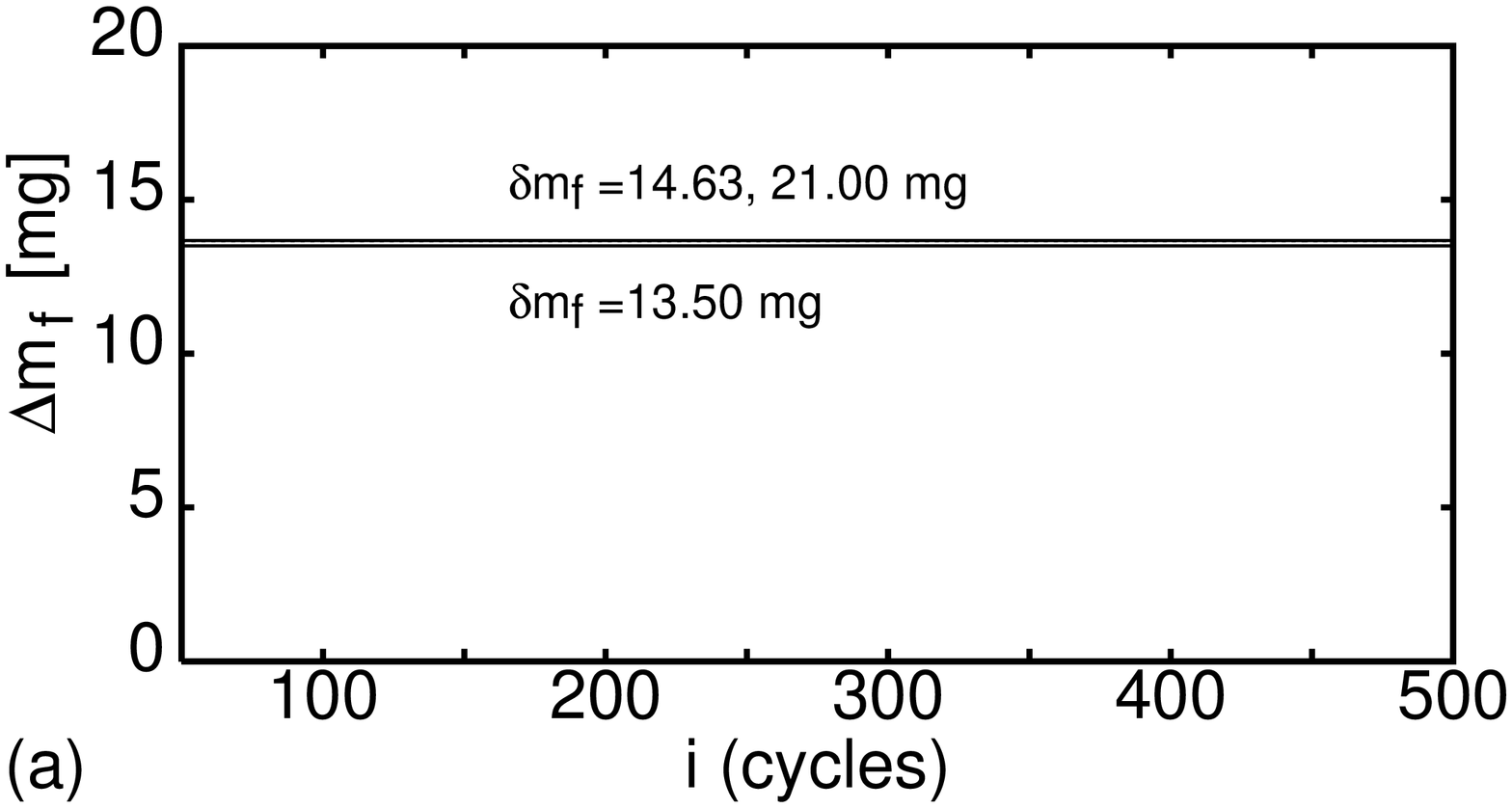}
\includegraphics[scale=0.3,angle=-90]{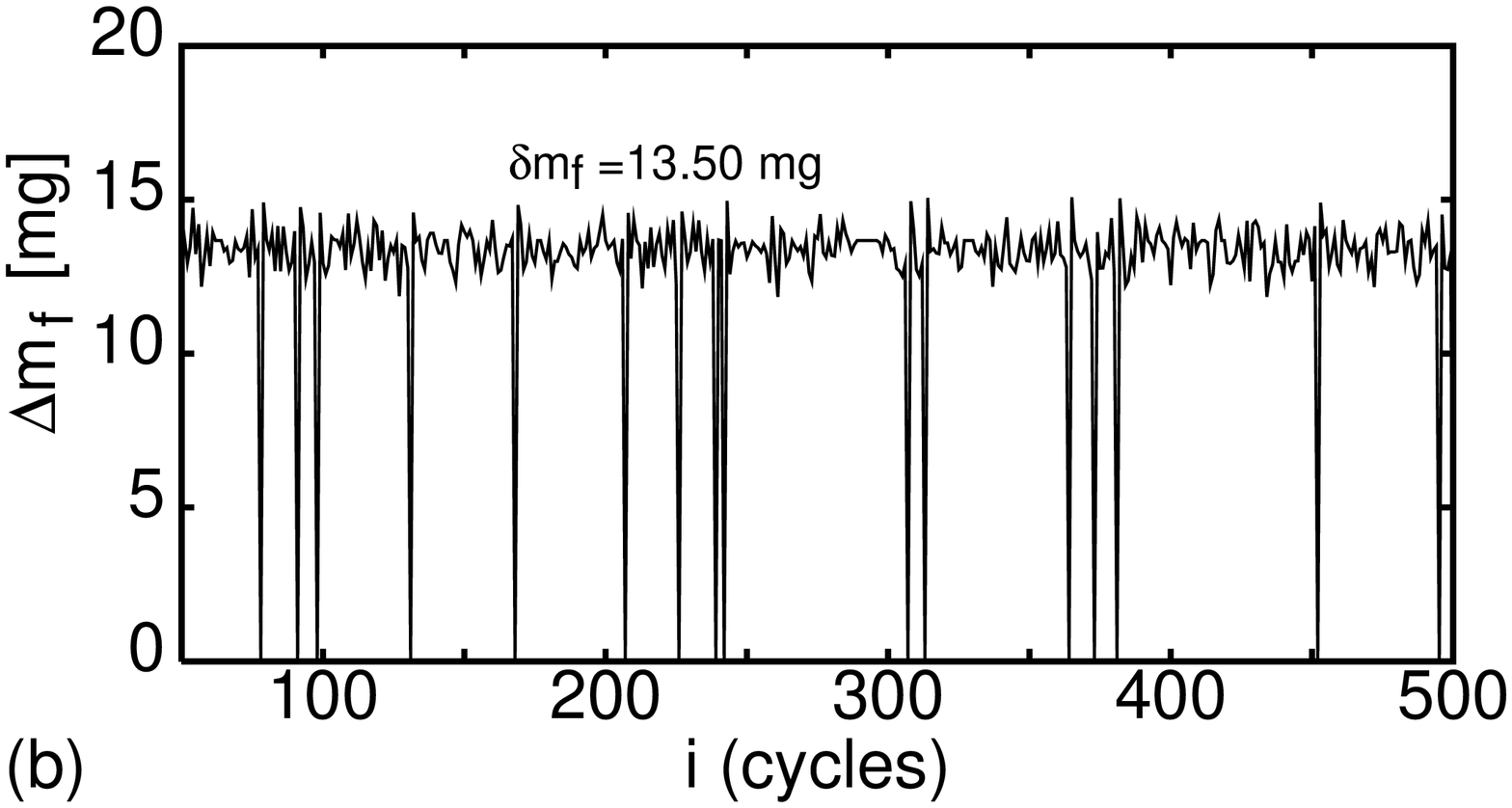}
\includegraphics[scale=0.3,angle=-90]{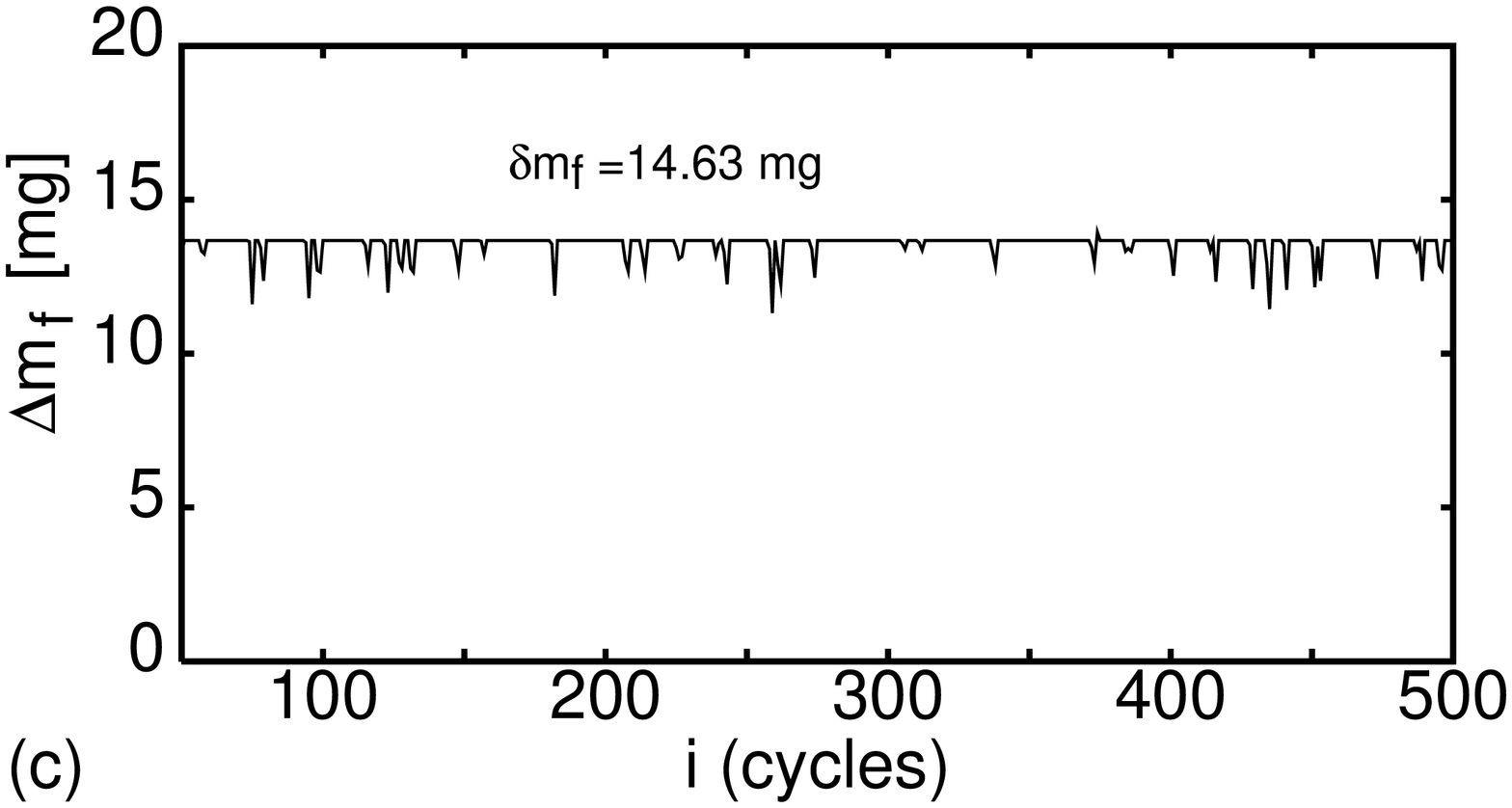}
\includegraphics[scale=0.3,angle=-90]{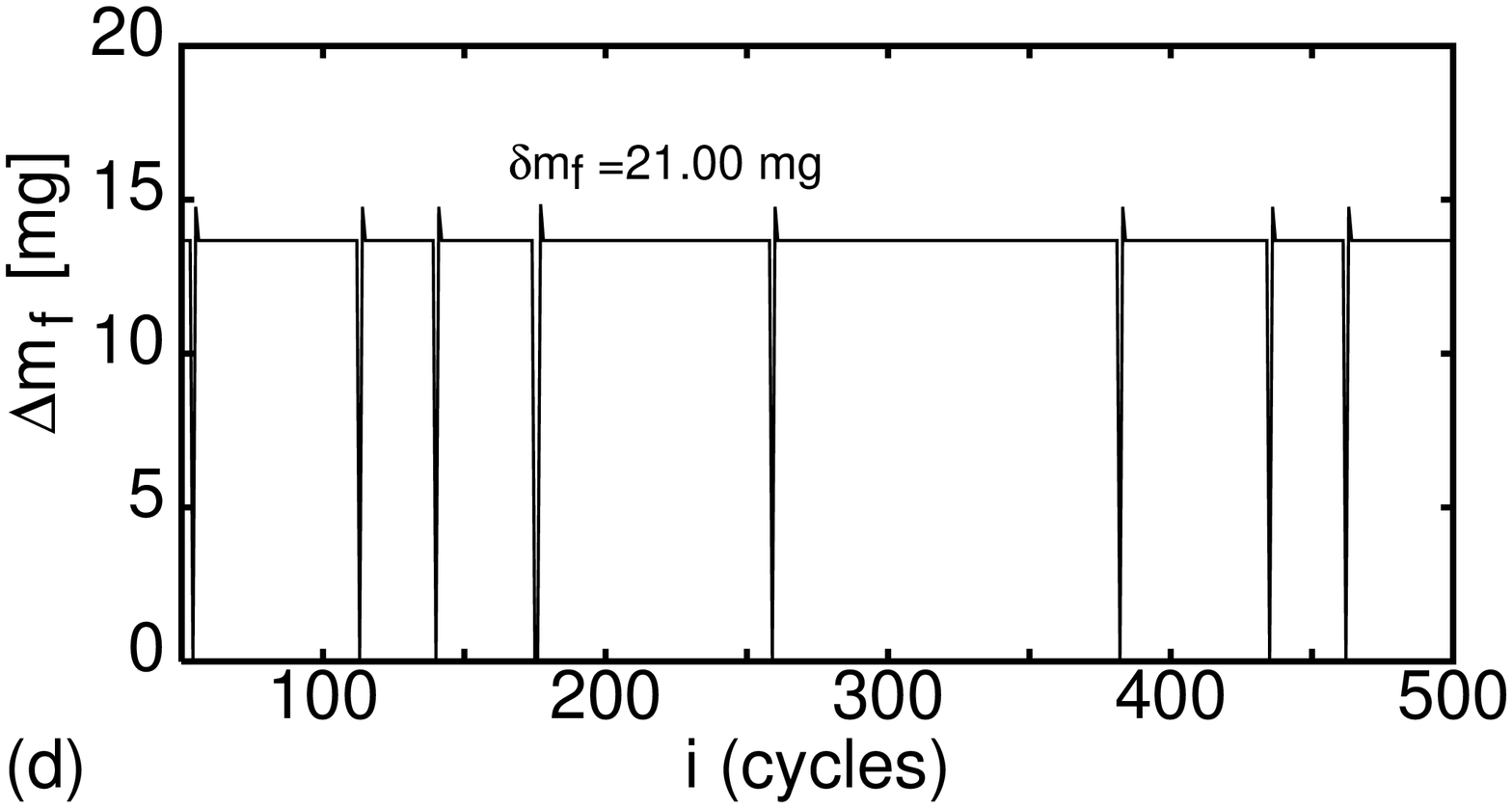}
\caption{The dependence of burned fuel mass 
on sequential cycles $i$ for deterministic (a) and
stochastic (b-d) processes.
$\delta m_a=200$ mg while
$\delta m_f$ takes different values: 13.50 mg , 14.63 mg, 21.00 mg denoted in
particular figures.}
 \end{center}
\end{figure} 

The results for burned fuel mass $\Delta m_f$ are presented in Fig. 3.
Starting from
deterministic conditions ($\delta m_{f}=\delta m_{fo}={\rm const.}$)
we obtain the constant fraction of the burned fuel mass $\Delta m_f$
represented by the three straight lines in Fig. 3a. 
All lines are lying
very close to
each
other and they are hardly distinguishable in Fig. 3a,
but actual numbers shows
clearly that $\Delta m_f$ increases
slightly with growing  $\delta m_f$
($\Delta m_f=13.50$ mg for $\delta m_f=13.50$ mg  while  $\Delta 
m_f=13.67$
mg for $\delta m_f=14.63$ mg and 21.00 mg) account for combustion 
constraints (Eqs. 1-7) and combustion curve (Fig. 1).

In Fig 3 b-c we show the same, $\Delta m_f$, for the considered case of
assumed
fuel injection ($\delta m_{fo}=13.50$ mg - Fig. 3b,
$\delta m_{fo}=14.63$ mg - Fig. 3c,
$\delta m_{fo}=21.00$ mg - Fig. 3d)
 and stochastic conditions. Due to different magnitudes parameter 
$\lambda$
fluctuations, and dependence of combustion curve Fig. 1 it is not
surprise that the fluctuations of $\Delta m_f$ have different character in
all these cases. For lean combustion, which is a stable process in
deterministic case, the fuel injection fluctuations introduce considerable
instabilities to the combustion process leading to the suppression of
combustion because in some cycles (Fig. 3b) where $\lambda$ is larger that
$1.3$. 

\begin{figure}
 \begin{center}
\includegraphics[scale=0.28,angle=-90]{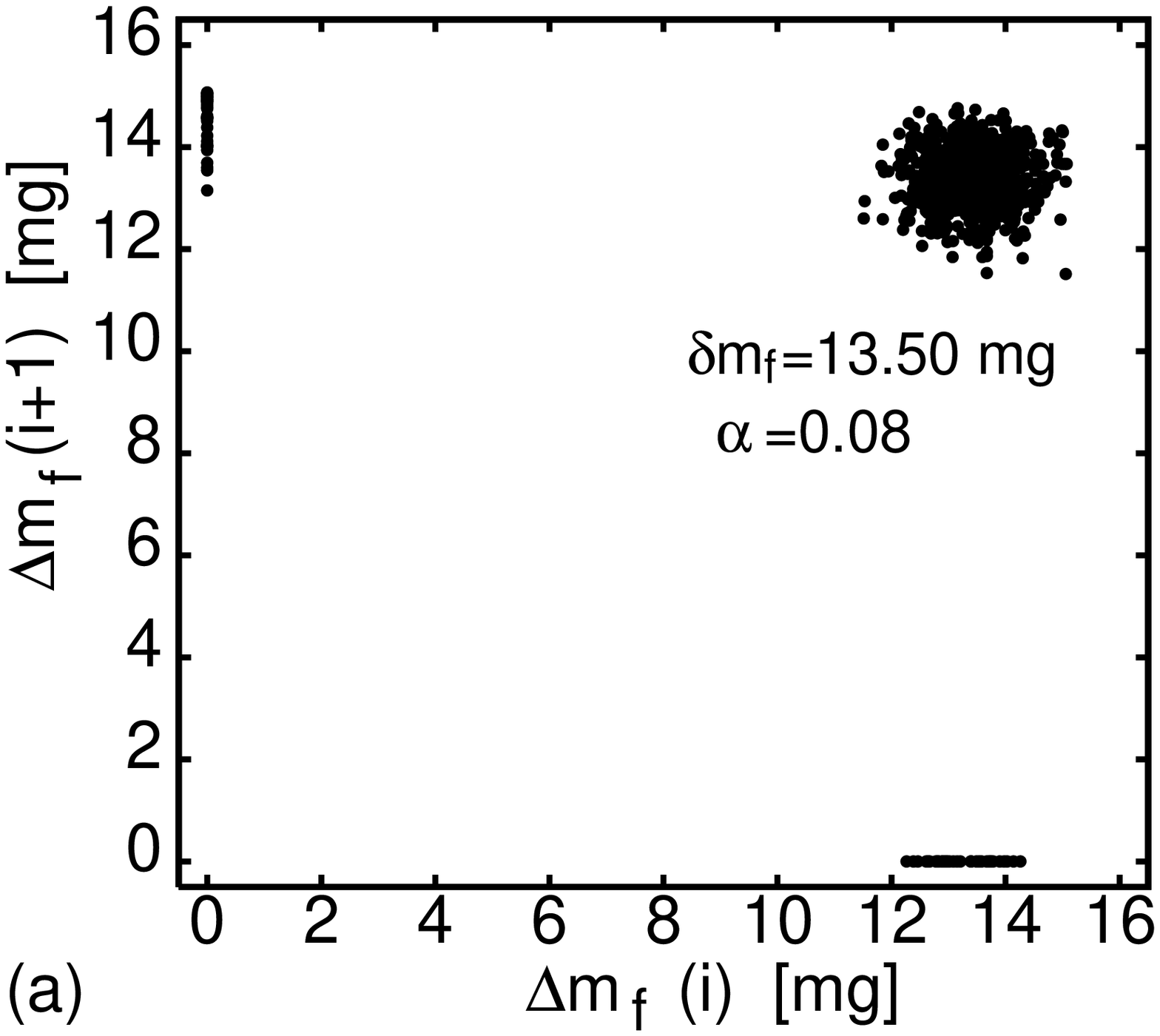} \hspace{-1.0cm}
\includegraphics[scale=0.28,angle=-90]{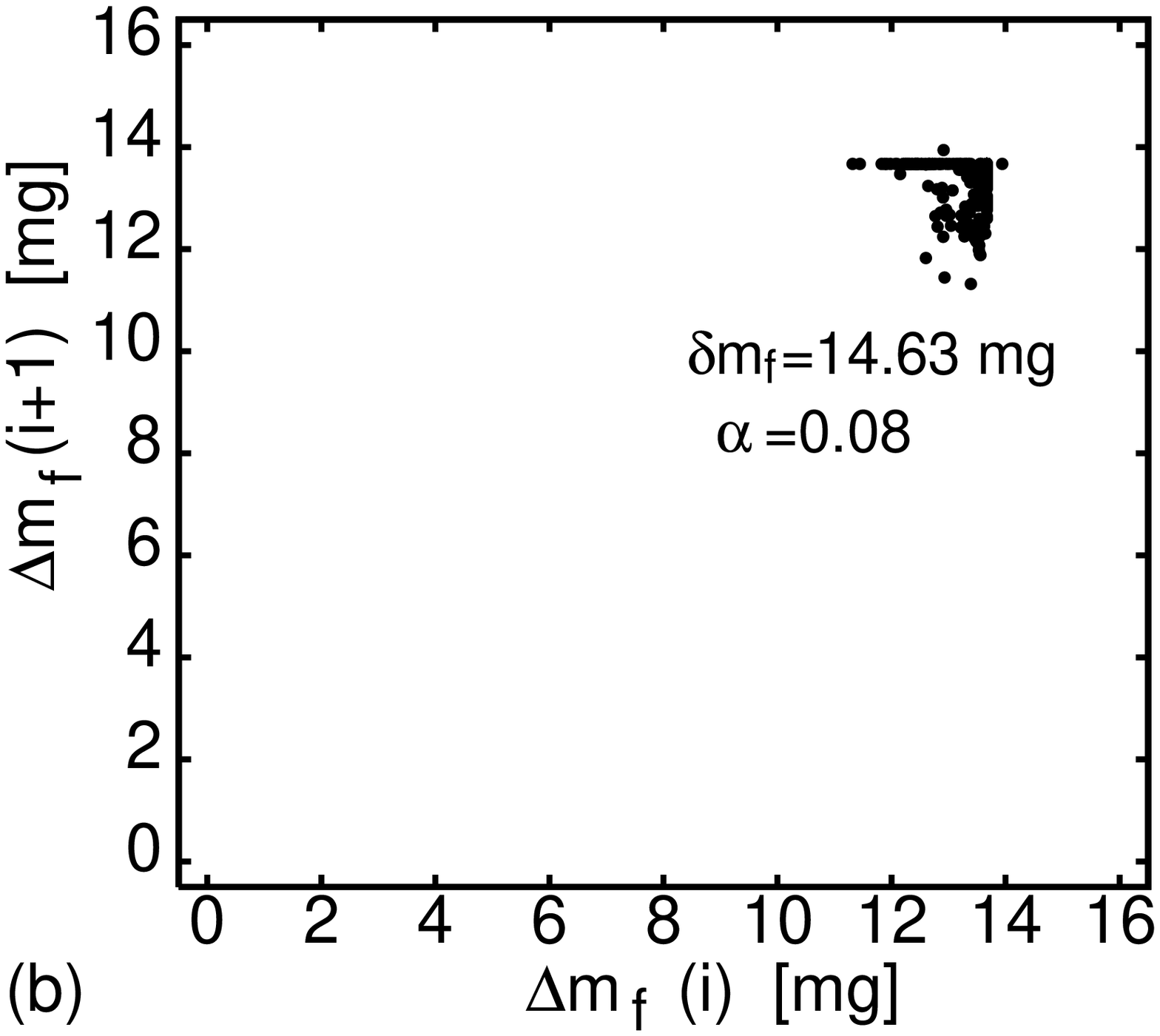} \hspace{-1.0cm}
\includegraphics[scale=0.28,angle=-90]{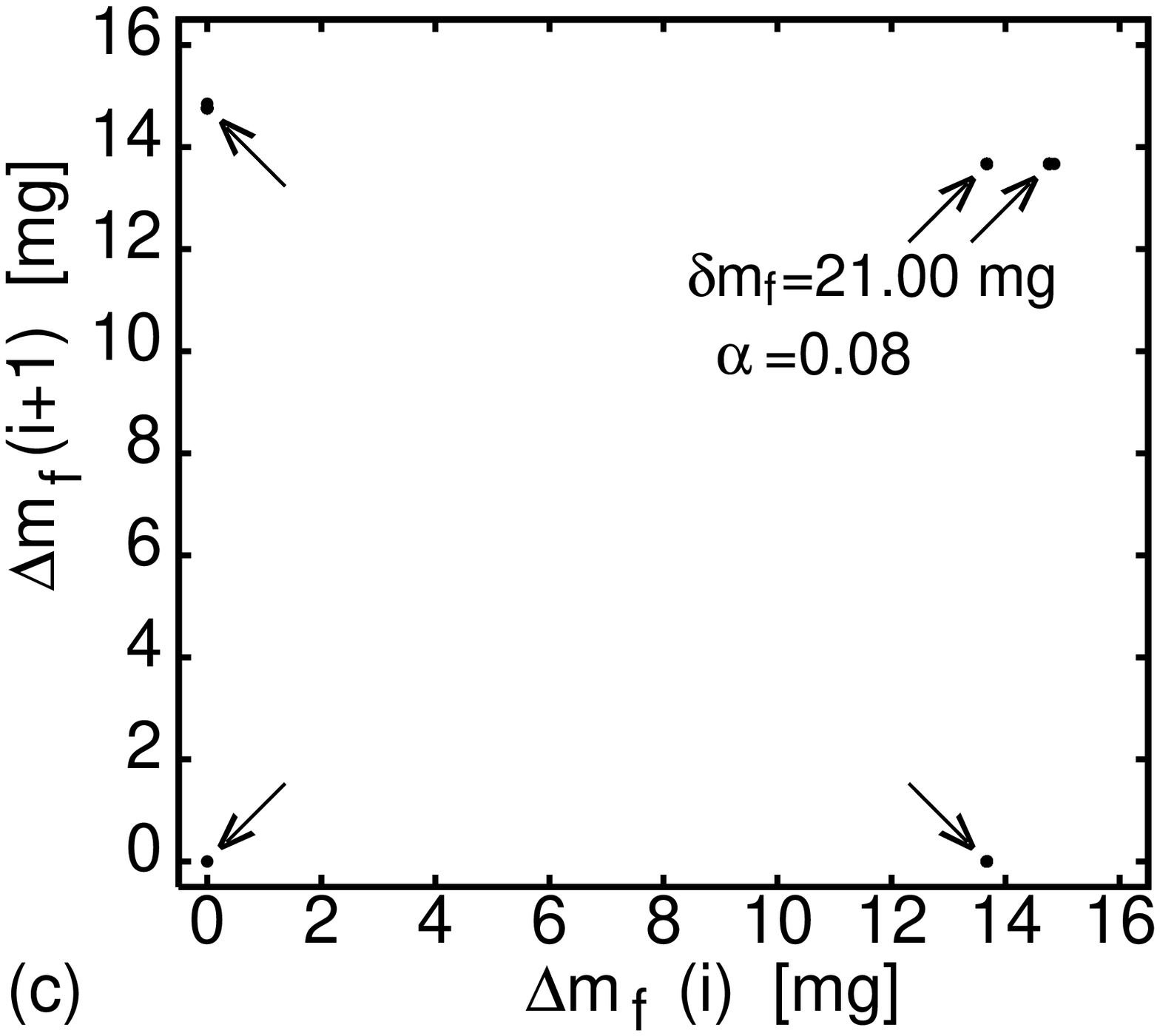}

\includegraphics[scale=0.28,angle=-90]{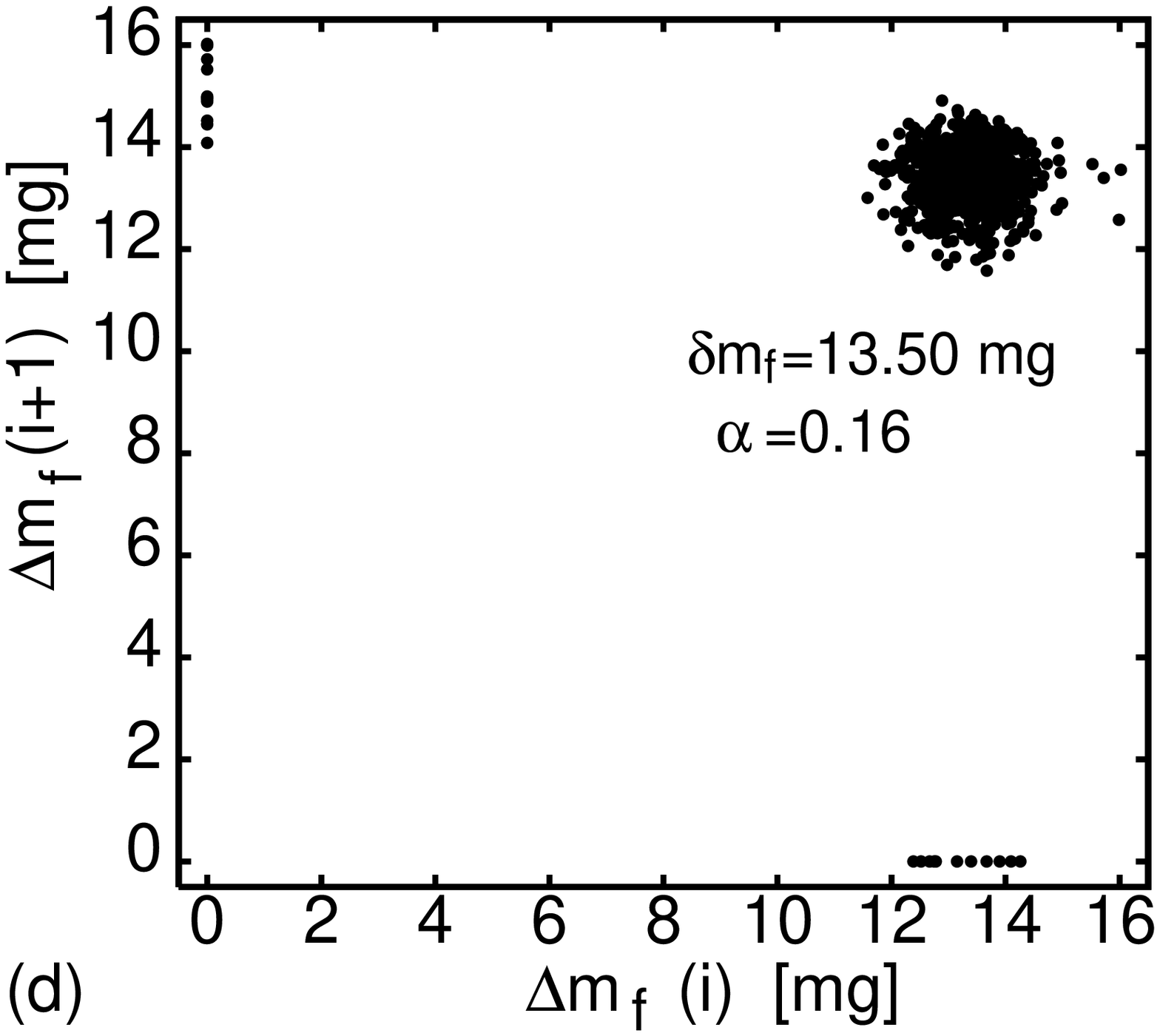} \hspace{-1.0cm}
\includegraphics[scale=0.28,angle=-90]{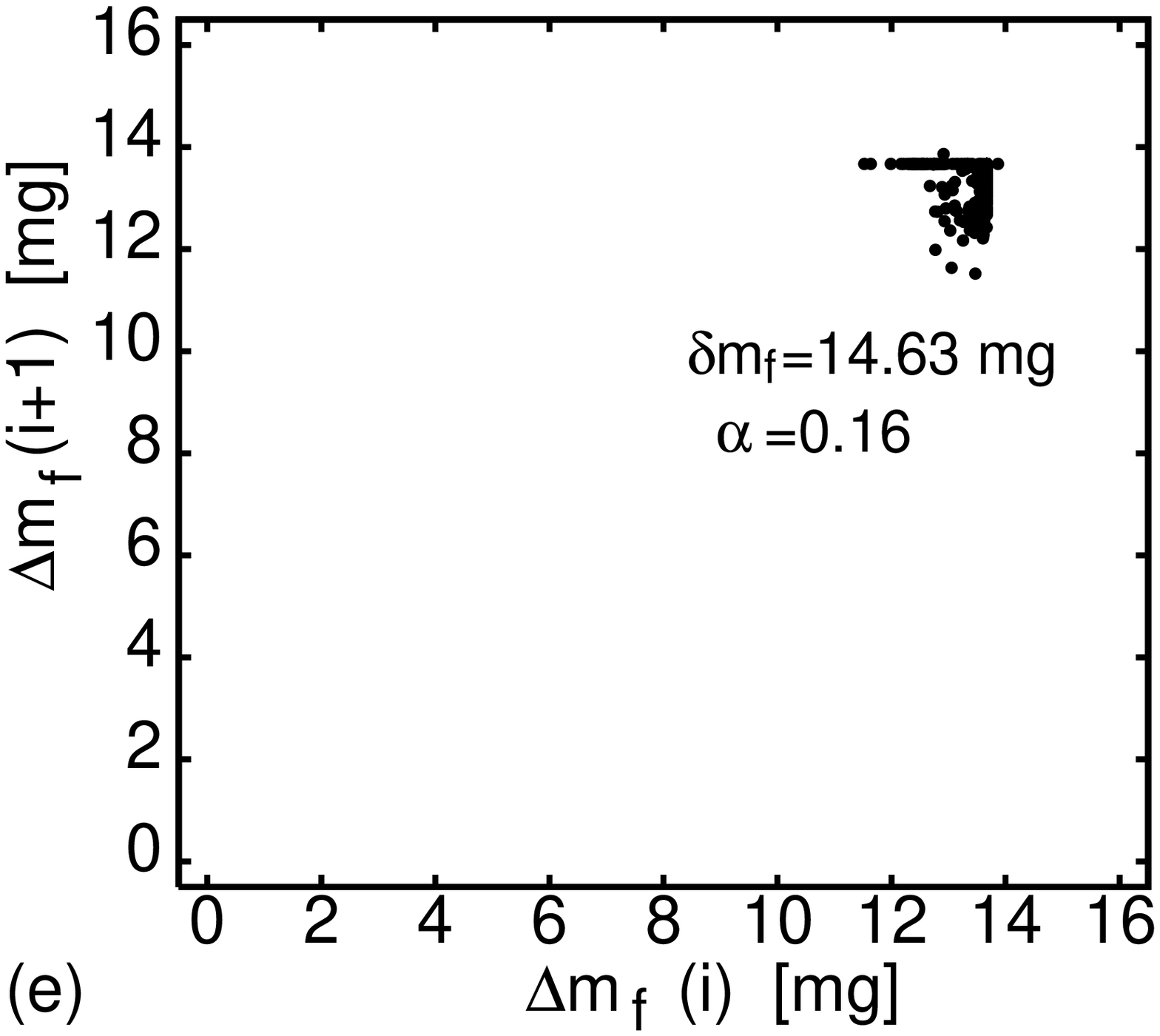} \hspace{-1.0cm}
\includegraphics[scale=0.28,angle=-90]{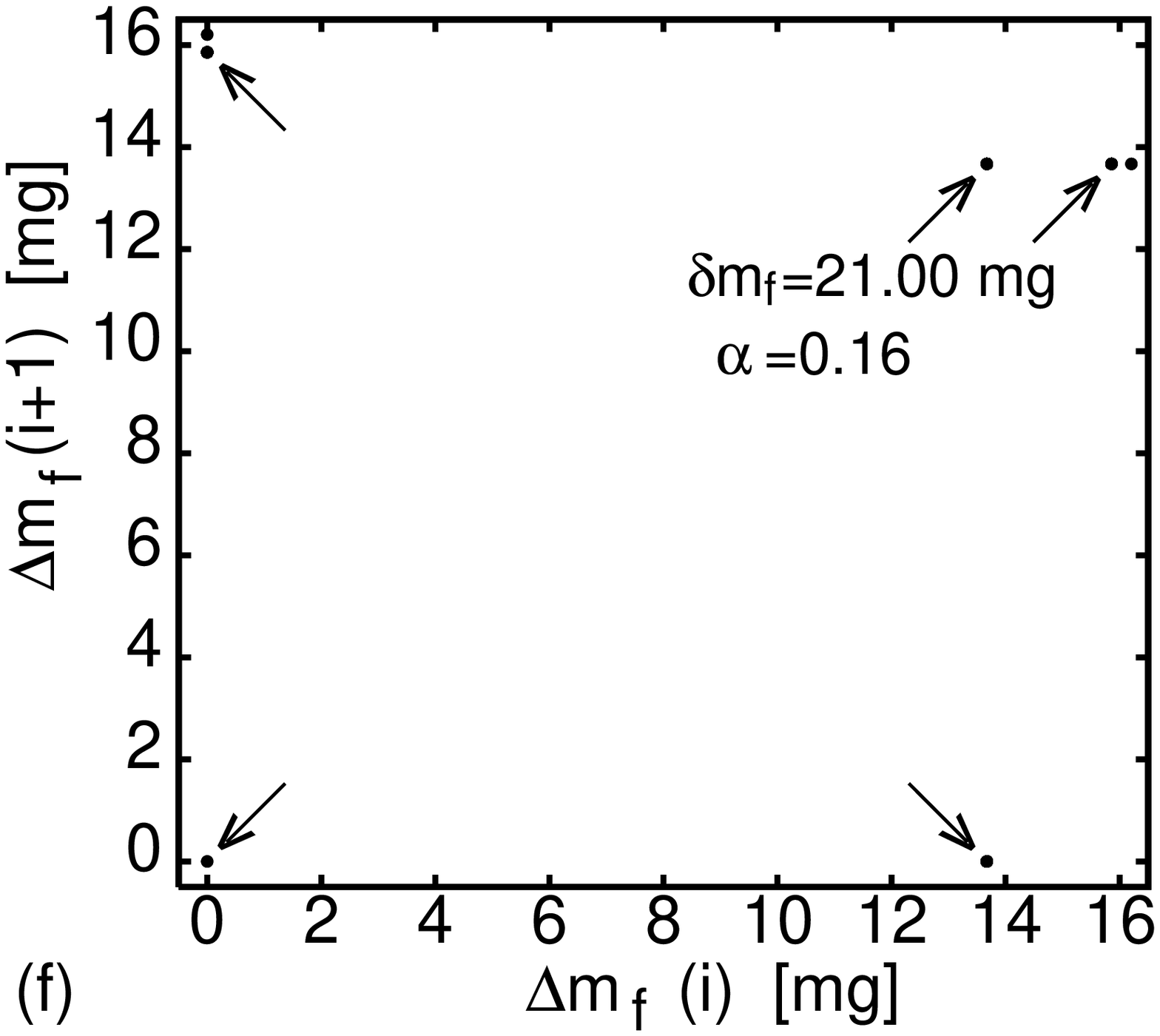}

\caption{Return maps for burned fuel mass: $m_i(i+1)$ versus $m_i(i)$, where $i$ denotes 
a present cycle 
for 
stochastic process ($\gamma=10\%$).
$\delta m_a=200$ mg while
$\delta m_f$ takes different values: 13.50 mg , 14.63 mg, 21.00 mg denoted 
in
particular figures.
Note, arrows in Figs. 4c and f indicate singular points on return maps for $\delta m_f$=21.00 mg.
}
 \end{center}
\end{figure}

\begin{figure}
\begin{center}
\includegraphics[scale=0.3,angle=-90]{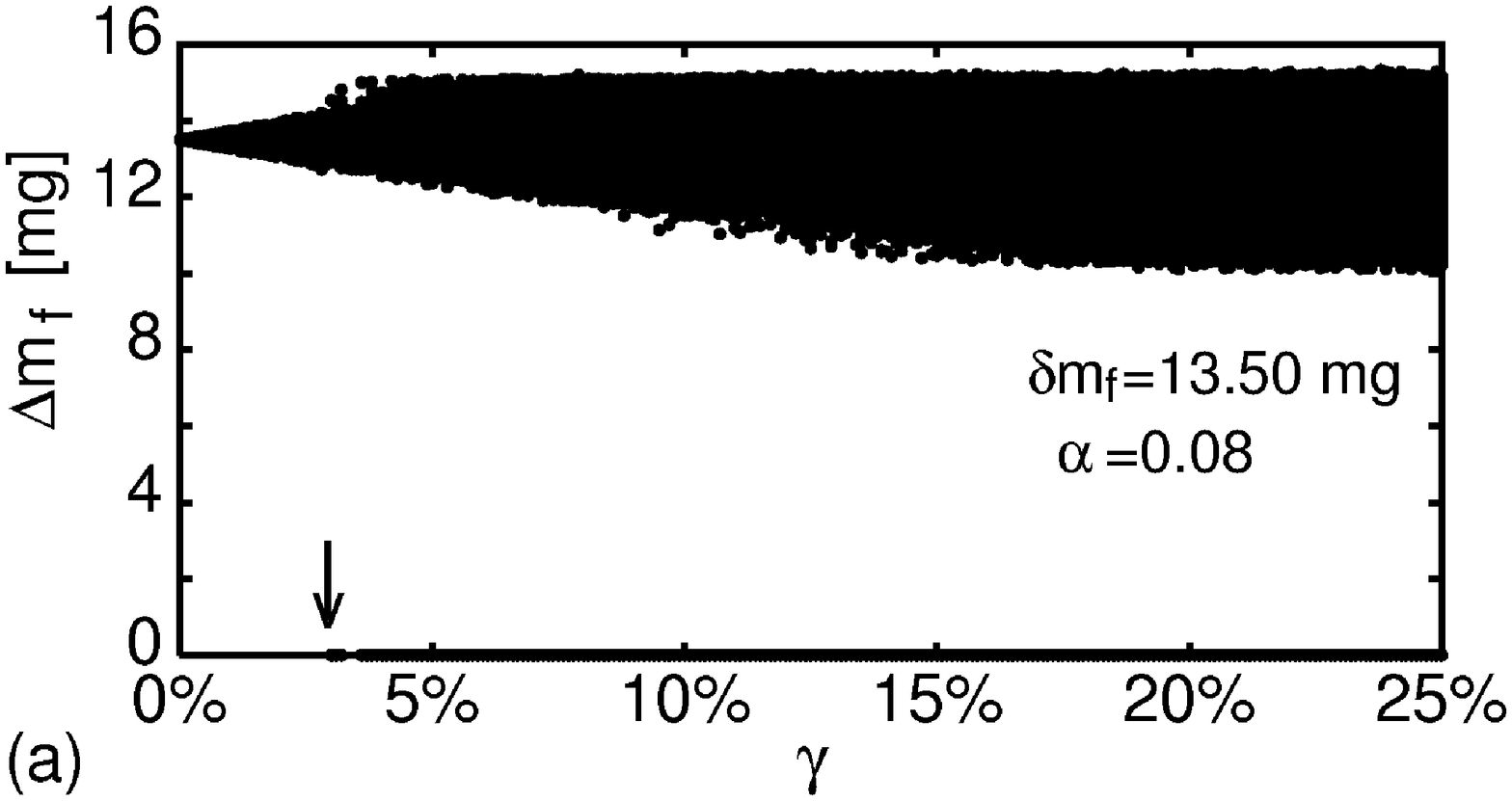}
\includegraphics[scale=0.3,angle=-90]{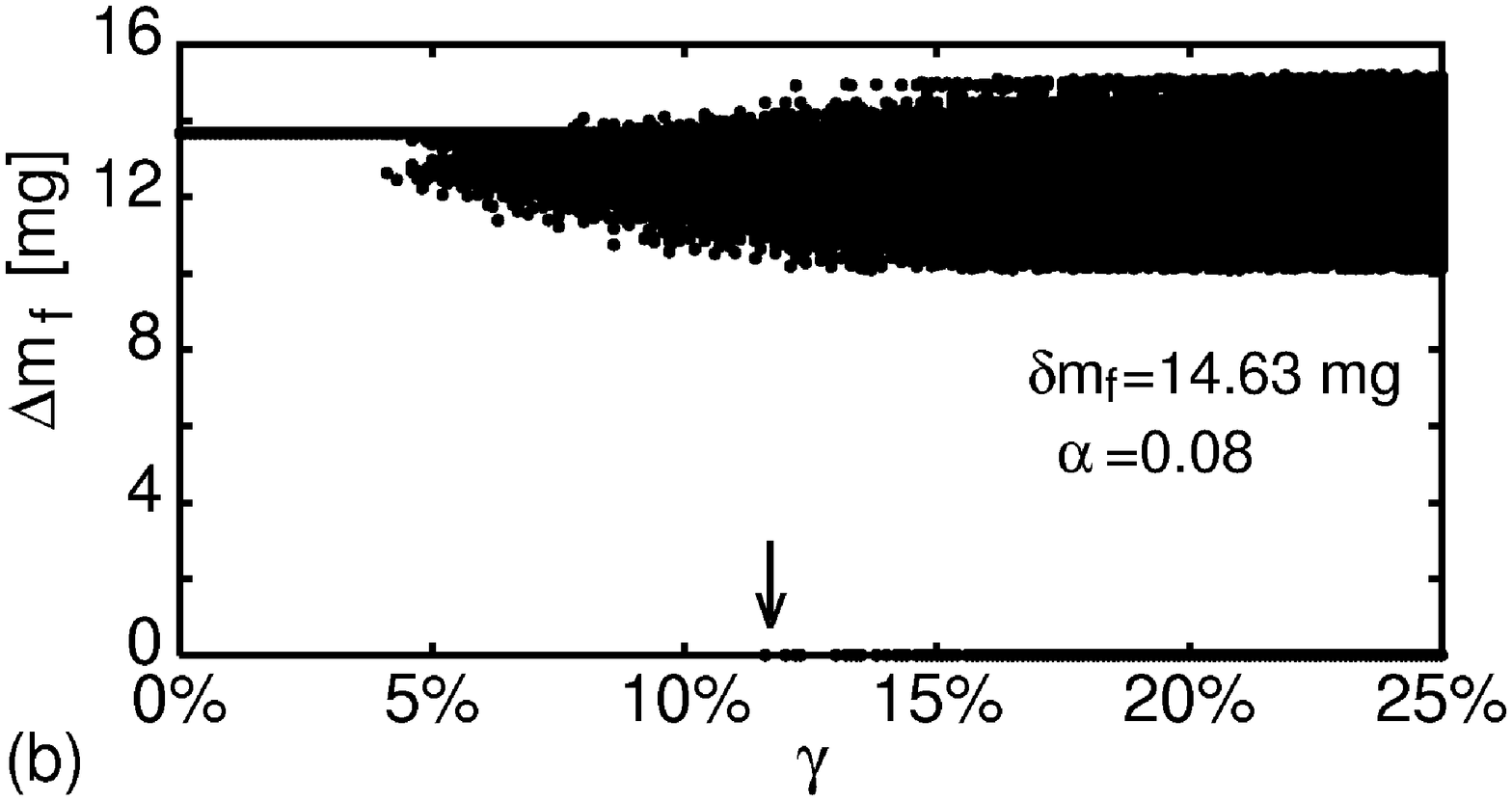}
\includegraphics[scale=0.3,angle=-90]{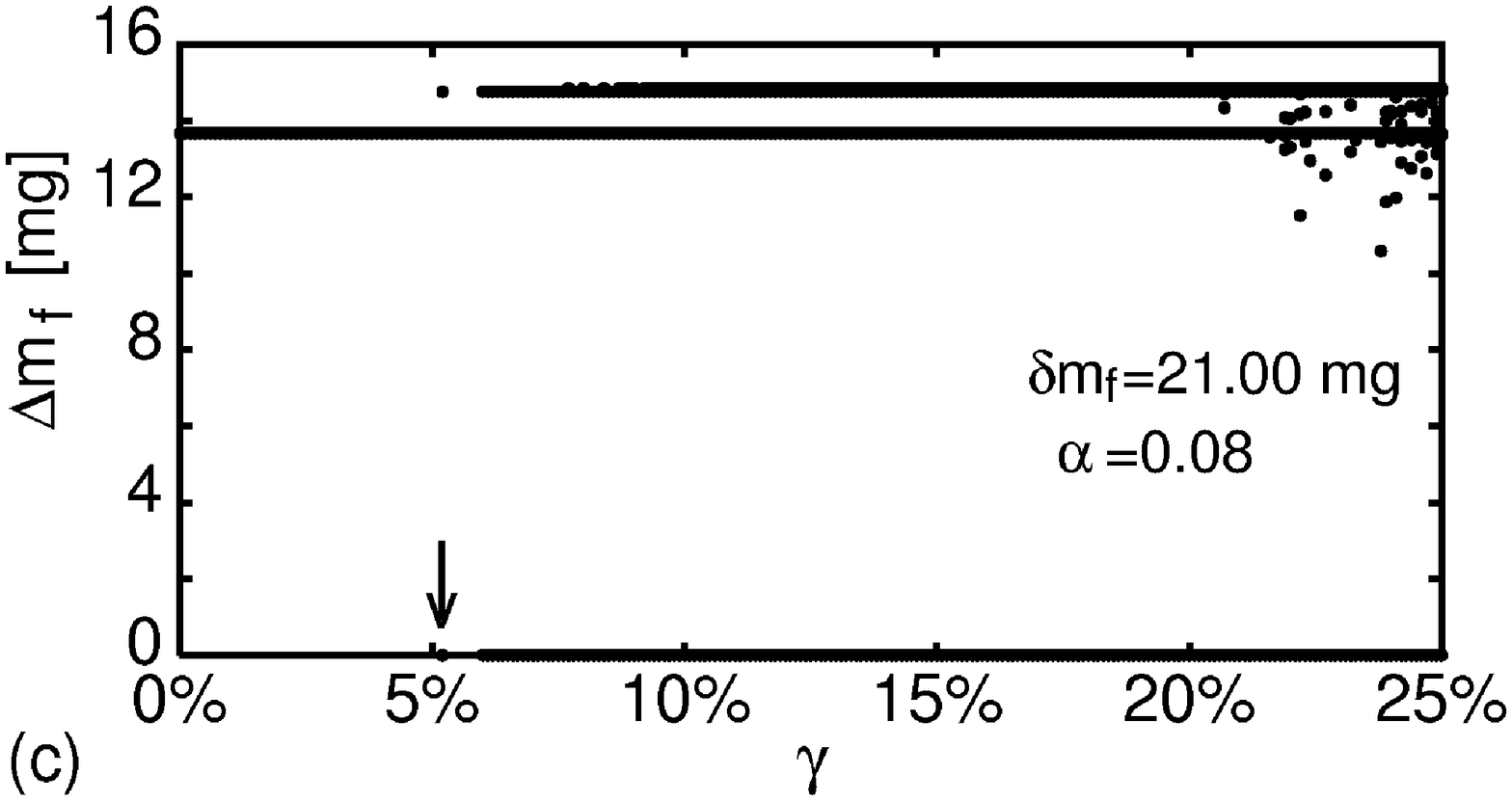}
\includegraphics[scale=0.3,angle=-90]{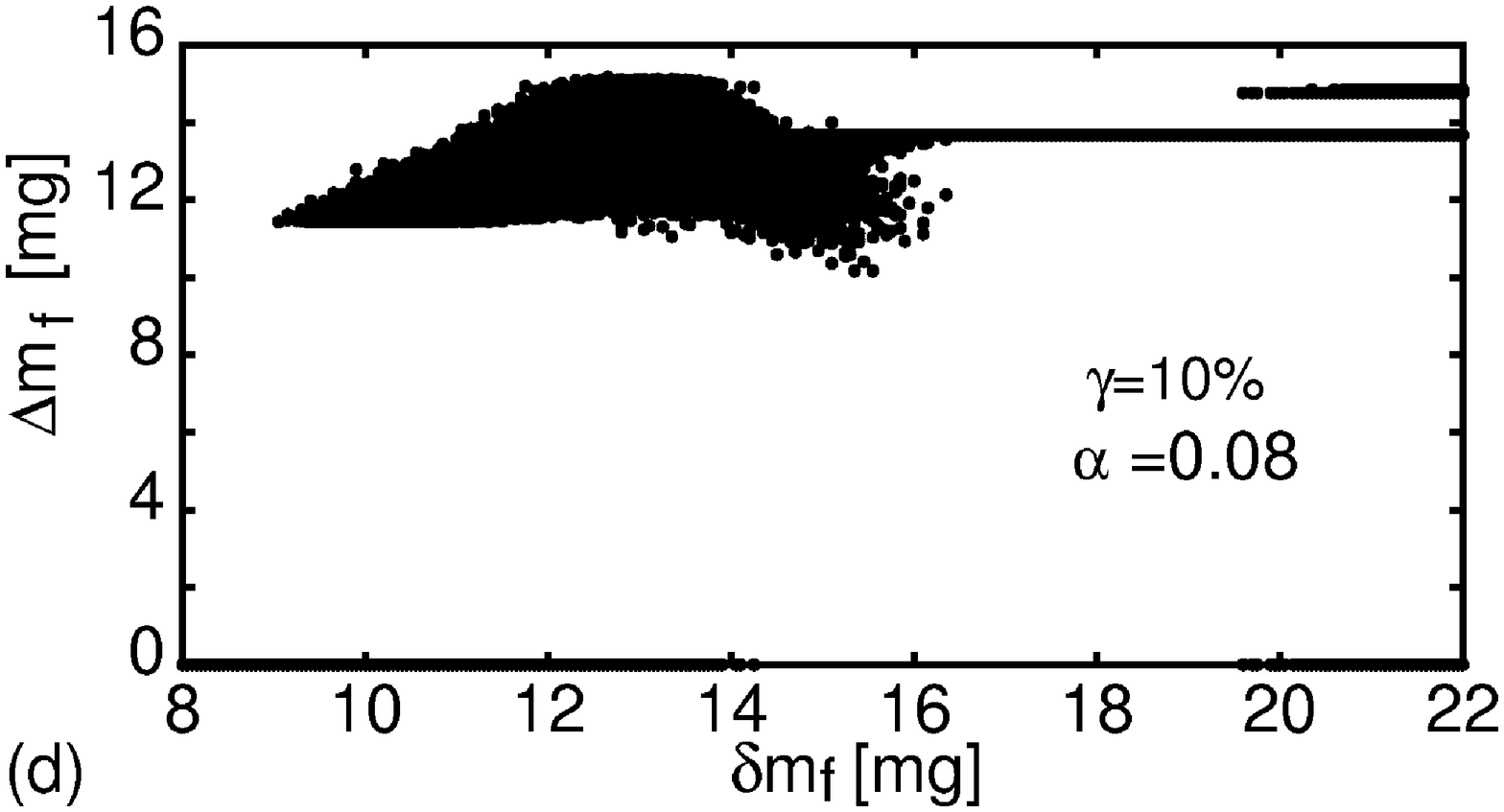}

\caption{Bifurcation diagrams.   $\delta m_f$ against noise (Fig. 5a-c) and against
the fresh fuel amount constant $\delta m_f$ for $\alpha=0.08$.
Arrows indicate a bifurcation regarding to misfires appearance.}
\end{center}
\end{figure}

\begin{figure}
\begin{center}
\includegraphics[scale=0.3,angle=-90]{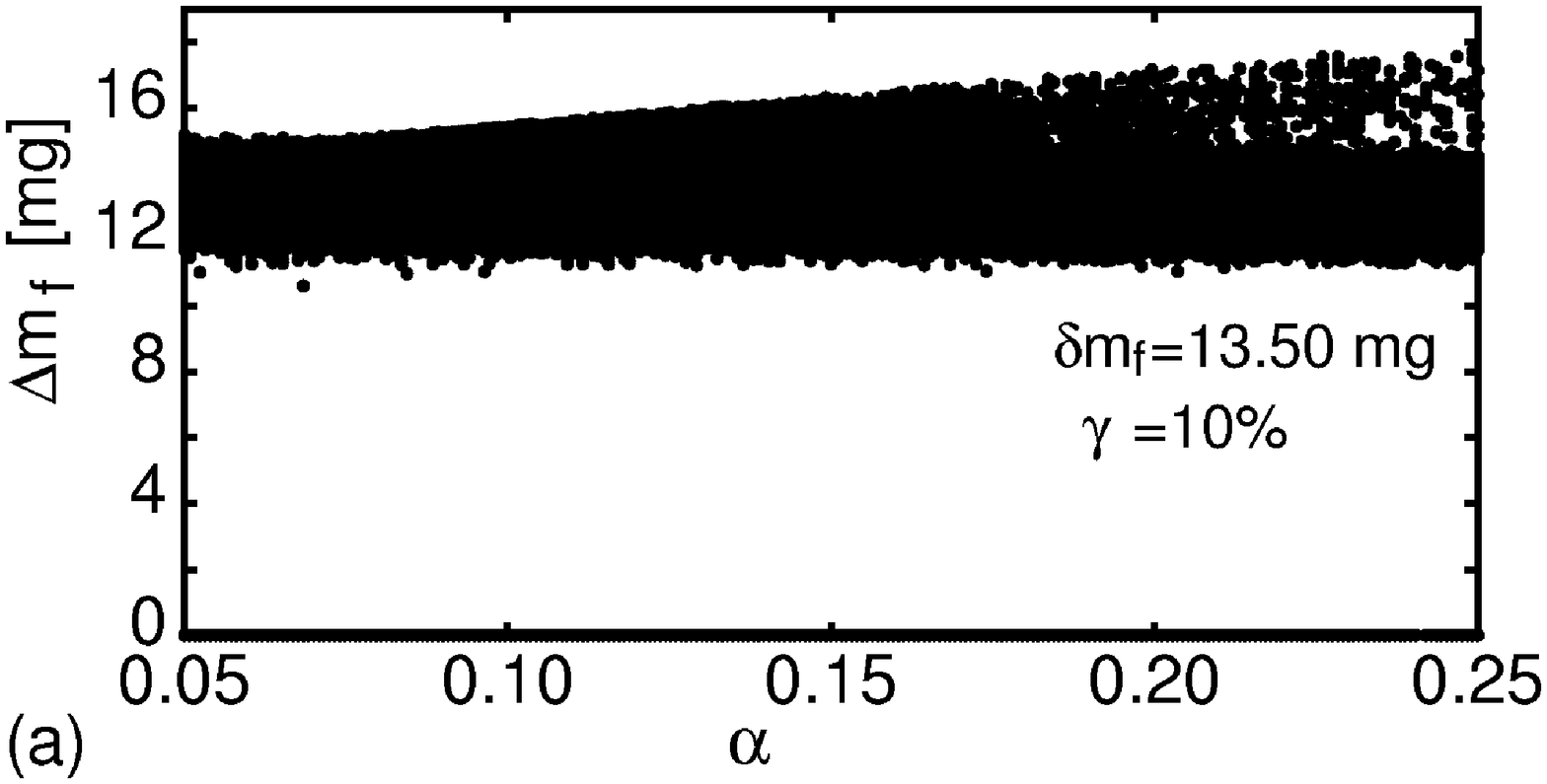} 
\includegraphics[scale=0.3,angle=-90]{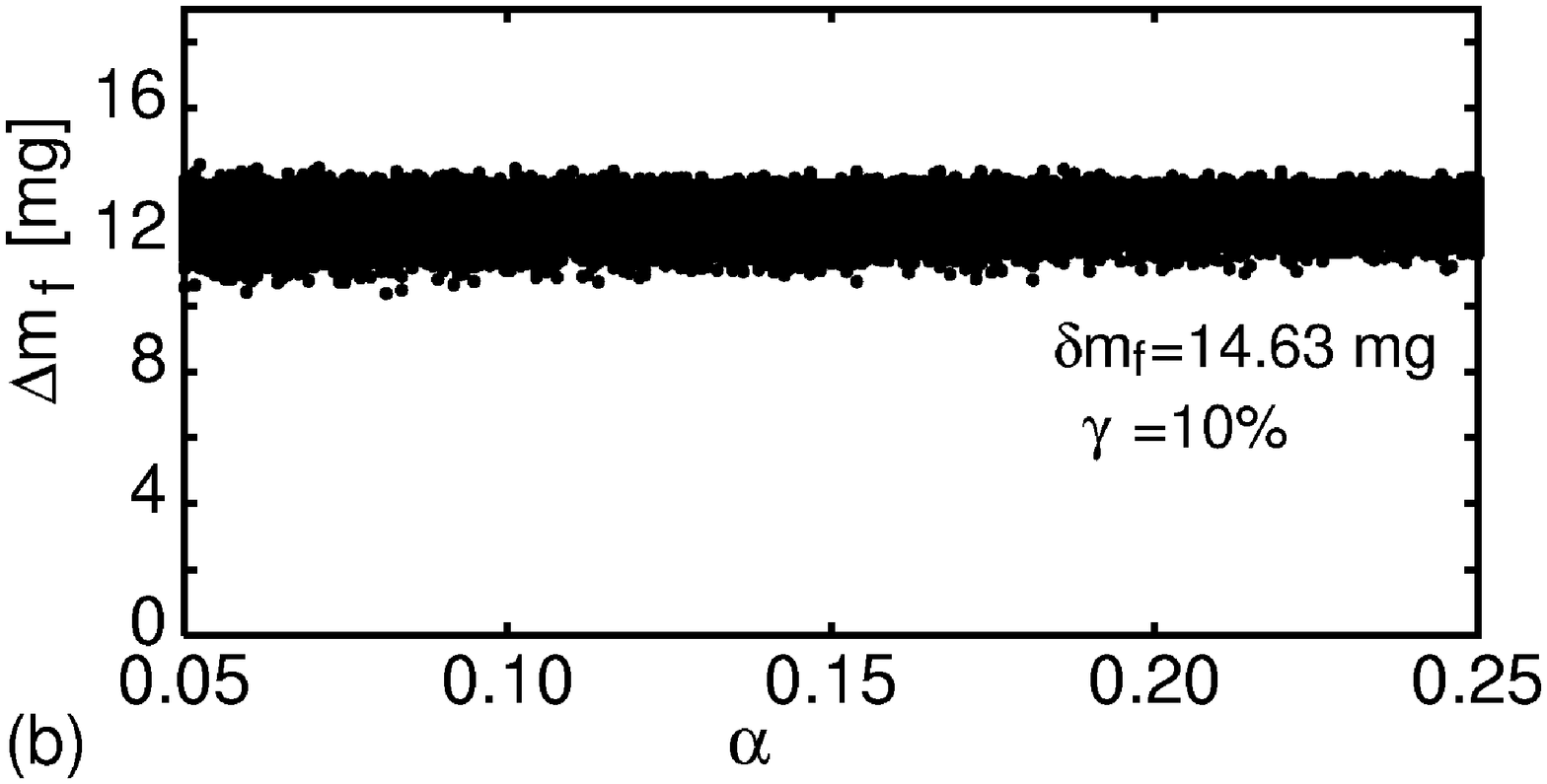}
\includegraphics[scale=0.3,angle=-90]{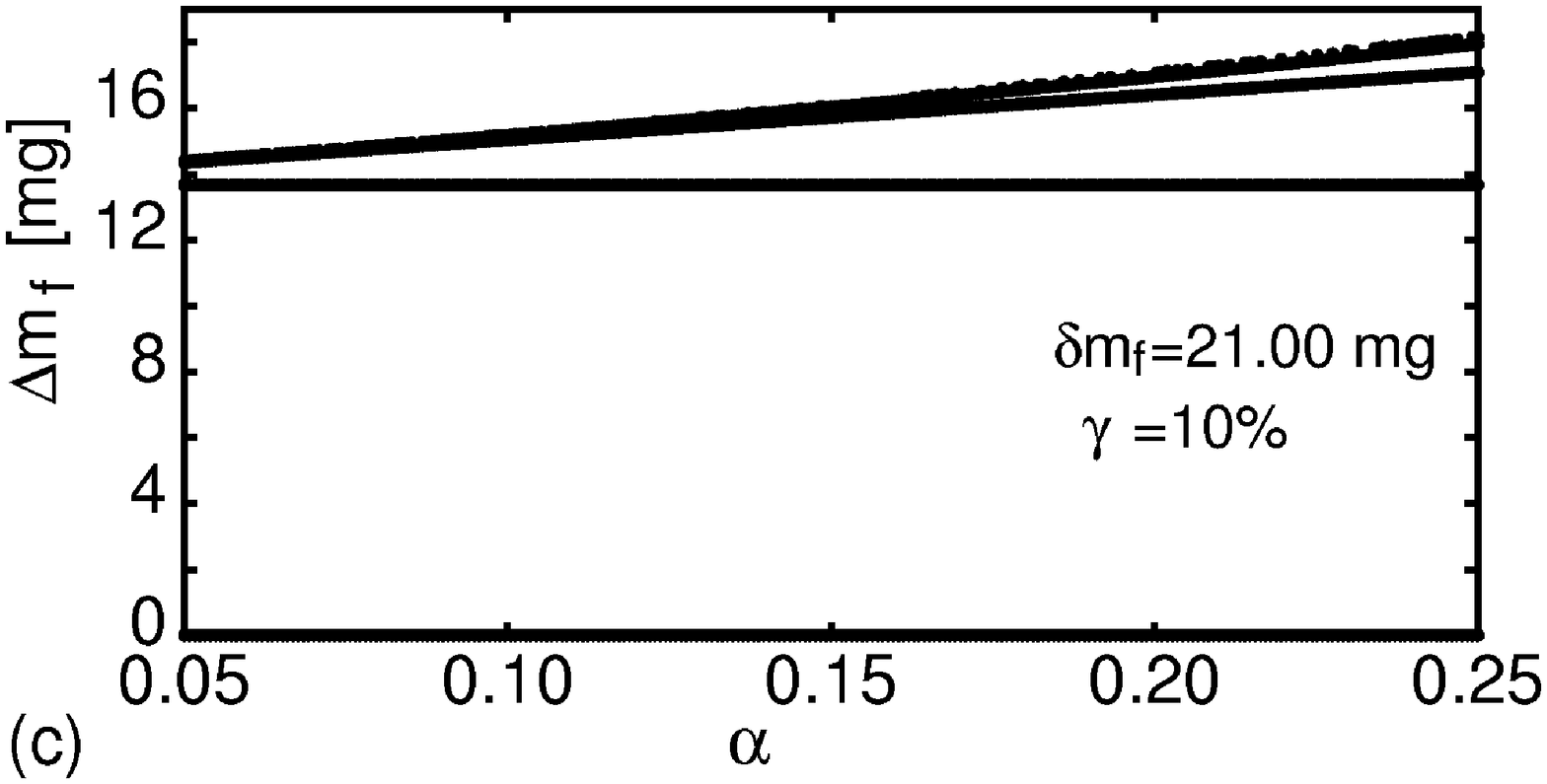}
   
\caption{Bifurcation diagrams:   $\Delta m_f$ against a residual gas 
fraction $\alpha$
for various fresh fuel amounts $\delta m_f$.}
\end{center}
\end{figure}

Then Equation  7 is not satisfied. In the next case  (Fig. 3c) 
the 
effect 
of stochasticity 
is much smaller. Here we have optimal air-fuel mixture. 
First of all one should note that fluctuations of 
$\lambda$ are smaller than in previous case (Fig. 2b). Moreover  
$\lambda$
oscillate around  
the region ($\lambda 
\approx 1$)  in
combustion curve 
(Fig. 1) which does not have big changes comparing to previous case.
Finally, Fig. 3d shows the sequence of $\Delta m_f$ for the large
$\delta m_{fo}$ (rich fuel-air mixture). The fluctuations of $\lambda$ 
are 
the smallest of all 
three ones but $\lambda \approx 0.7$ causes suppressions of combustions
 in some cycles similarly to the case shown in Fig. 3b.

In Figs. 4a-c we have plotted return maps of all
considered stochastic cases Figs. 3b-d. Additionally, we have shown similar plots for higher residual gas fraction 
$\alpha=0.16$ (Figs. 4d-f). Note that differences between both cases ($\alpha$ =0.08 and 0.16) are small.
It is also worth to note this results are 
agreement with diagrams of experimental heat release data for engines {\em Kohler} and {\em Quad4}
presented by Green Jr. and coworkers (Fig. 1 in  Green et al. 1999) for three
values of an equivalence ration. Of course heat release
can be treated, in the first approximation,  as proportional to burned mass through a heating constant of 
fuel. The only 
important difference  
visible between our simulated and experimental data is that the second is more smeared. This could be connected 
with our model assumptions basing on two components. In reality there are more  components
(Heywood 1988, Rocha-Martinez {\em et al.} 2002)  and 
consequently instead of the simple relation between 
burned mass and  heat release there is  more sophisticated one. The second, more important, reason may lie in 
our assumed combustion characteristics (Fig. 1).  It eliminates partial combustion in the initial stage and 
slow development 
of flame  for smaller $\lambda$ may be important for combustion process dynamics (Hu 1997).    
Our results are also close to that obtained from simulations by Daw {\em et al.} (Daw {\em et al.} 2000,
Fig. 9).

For better clarity we have also plotted bifurcation diagrams with respect to stochastic parameter 
$\gamma \in [0\%, 22\%]$ (Figs. 5a-c) as well as  a fresh fuel feeding constant $\delta m_f \in [8,22] $
(Figs. 5d)  and  
$\gamma=10\%$. Note, single points in vertical cross sections indicate stable combustion,
multiple points combustion instability with misfire and possible oscillations. Finally, 
dark regions identify combustion with stochastic oscillations and possible intermittent misfire. 

In Figs. 6a-c we have shown bifurcation diagrams against 
residual gas fraction $\alpha$ ($\alpha \in [0.05, 0.25]$). It is clear 
that
this parameter has a small influence on combustion fluctuations. 
We have not observed any qualitative changes
for all considered fresh fuel amounts $\delta m_f$.

\section{Conclusions}

In this paper we examined the origin of burned mass fluctuations in a 
simple model of combustion.
In case of stochastic conditions we have shown that depending on the 
quality of fuel-air mixture the final effect is different. The worse
situation is for lean combustion. The consequences of it can be observed
for idle speed regime of engine work. Unstable engine work, interrupted by
the  cycles with misfire, leads to a large increase of fuel use.

Although the presented two component model is very simple it can reflect
the underlying nature of engine working conditions. In spite of fact 
that the model is characterised by the nonlinear transform 
(Eqs. 3, 5 and 7) similar to logistic one, we have not 
found 
any chaotic region. On the other hand  we do find qualitative and quantitative features given  in the earlier
experimental works
 (Green Jr. {\em et al.} 1999).
However we cannot exclude  that chaotic solutions can be found for other 
engine parameters.

The other strong limitation was 
concerned with the sharp edges of combustion curve
($\Delta m_f$ versus $\lambda$ in Fig.1) modelled by a sharp decay (step function). We used such an approximation as  
 a simplest one  

It can be improved but modelling with the exponential growth $\exp(-1/x)$.
Such assumption would be  more 
realistic as it would correspond to partial combustion where 
the mixture of air and fuel is changing its properties inside the 
cylinder. 
From a physical point of view mixture
gasoline-air is
not uniform before ignition and that can cause nonuniform combustion
smearing the edges of the combustion curve (Fig. 1).
In such a case slow development of early flame can also vary from time to time.
Going in that direction one can also include additional 
dimensions to incorporate  diffusion phenomenon (Abel {\em et al.} 2001).

However assumptions about the exponential dependence of combustion curve 
may involve an averaged  effect.  We think that it  led to 
chaotic behaviour in earlier papers  (Daw {\em et al.} 
1996, 1998, Green Jr. {\em et al.} 1999).
Calculations considering this effect in our model are in progress and the results will 
be reported in 
a separate future publication. 
\vspace{1cm}

\noindent {\em Acknowledgement. The authors are very grateful to unknown referees for their constructive 
remarks. GL would like to thank International Centre for Theoretical Physics in  Trieste for hospitality.}

\section*{References}

\noindent
Abel, M., Celani, A., Vergni, D., and Vulpiani, A., 2001, "Front propagation in laminar flows"
{\em Physical Review} E {\bf 64} 046307.
\\ \\
\noindent 
Clerk D., 1886,  {\em The gas engine}, Longmans, Green \& Co.,
London.
\\ \\
Chew L., Hoekstra R., Nayfeh, J.F., Navedo J., 1994, "Chaos analysis of
in-cylinder pressure measurements", {\em SAE Paper}  942486. 
\\ \\
Daily J.W., 1988, "Cycle-to-cycle variations: a chaotic process ?"
{\em Combustion
Science and Technology} {\bf 57}, 149--162.
\\ \\
Daw, C.S., Finney, C.E.A., Green Jr., J.B.,
Kennel,
M.B.,
Thomas, J.F. and
Connolly, F.T., 1996, 
"A simple model for cyclic variations in a spark-ignition engine", {\em 
SAE
Paper}   962086. 
\\ \\
Daw, C.S., Kennel, M.B., Finney 
C.E.A. and  Connolly F.T., 1998
"Observing
and
modelling dynamics in an
internal combustion engine", {\em Physical  Review} E {\bf 57},
2811--2819. 
\\ \\
Daw, C.S., Finney, C.E.A. and Kennel, M.B., 2000, "Symbolic
approach for 
measuring temporal "irreversibility",  {\em Physical  Review} E,
{\bf 62},  1912--1921.
\\ \\
Foakes, A.P., Pollard, D.G., 1993, Investigation of a chaotic mechanism 
for
cycle-to-cycle variations, {\em Combustion Science and Technology} {\bf 
90},
281--287.  
\\ \\
Green Jr., J.B., Daw, C.S., Armfield, J.S., Finney, C.E.A., Wagner, 
R.M., Drallmeier, J.A.,
Kennel, M.B., Durbetaki, P., 1999, "Time irreversibility and comparison of
cyclic-variability models". {\em SAE Paper}   1999-01-0221.
\\ \\
Heywood JB., 1988,  {\em Internal combustion engine 
fundamentals},  McGraw-Hill, New
York. 
\\ \\
Hu Z., 1996, "Nonliner instabilities of 
combustion 
processes and
cycle-to-cycle variations
in spark-ignition engines", {\em SAE Paper}  961197. 
\\ \\
Kaminski T., Wendeker M., Urbanowicz K. and
Litak G., 2004, "Combustion process  in a spark ignition
engine: dynamics and noise level estimation", {\em Chaos} {\bf 14}  2 
(in 
press).
\\ \\
Kowalewicz A., 1984,  "Combustion systems of high-speed piston
i.c. engines", in {\em Studies in Mechanical Science} {\bf 3}, Elsevier, 
Amsterdam. 
\\ \\
Letellier, C., Meunier-Guttin-Cluzel, S., Gouesbet, G., Neveu, F., Duverger, T. and Cousyn, B., 1997, 
"Use of the nonlinear dynamical system theory to study cycle-to-cycle Variations from spark ignition engine 
pressure data", {\em SAE Paper} 971640.  
\\ \\
Roberts, J.B., Peyton-Jones J.C. and Landsborough
K.J., 1997, "Cylinder
pressure variations as a
stochastic process", {\em SAE Paper}  970059. 
\\ \\
Rocha-Martinez, J.A., Navarrete-Gonz\'alez, T.D., Pavia-Miller, C.G., 
P\'aez-Hern\'andez, R., 2002, "Otto and Disel engine models with cyclic 
variability", {\em Revista Mexicana De Fisica} {\bf 48}, 228-234.   
\\ \\
Shen, H., Hinze, P.C., Heywood, J.B., "A study of cycle-to-cycle variations in si engines using a modified 
quasi-dimensional model, {\em SAE Paper} 961187  
\\ \\
Wendeker, M.,  Niewczas, A. and Hawryluk, B., 2000, "A
stochastic model of
the fuel injection of the
si engine",  {\em SAE Paper}  2000-01-1088. 
\\ \\
Wendeker, M., Czarnigowski, J., Litak, G. and  
Szabelski, K., 2003, "Chaotic
combustion
in spark ignition engines", {\em Chaos, Solitons \& Fractals} {\bf 18}, 
805--808. 
\\ \\
Wendeker, M., Litak, G., Czarnigowski, J., and 
Szabelski, K., 2004, "Nonperiodic oscillations 
of pressure in a spark ignition
engine",  {\em International Journal of  Bifurcation and Chaos}  {\bf 14}, 
 5 (in press).

\end{document}